\documentstyle[fleqn,prb,aps,amssymb,preprint]{revtex}

\begin{document}
\draft
\widetext

\title{Model for the thermal conductivity of a quasicrystalline alloy}
\author{A. Jagannathan}
\address{
Laboratoire de Physique des Solides,
Universit\'{e} Paris--Sud,
91405 Orsay,
France }
\maketitle

\begin{abstract}
We present a simple model for the observed temperature dependence of the
thermal
conductivity of AlPdMn (Chernikov et al, Phys.Rev.B vol.51, (1995),153). We
 account for the low temperature data, upto and
including the plateau of the thermal conductivity, and discuss the
similarities and
the differences between the quasicrystal and amorphous solids.
\end{abstract}

\narrowtext
A recent paper by Chernikov, Bianchi and Ott \cite{cbo} presents thermal
conductivity data for the quasicrystal AlPdMn. The data are obtained for
excellent quality samples of $\rm Al_{70}Pd_{21}Mn_9$, containing very few
phason defects and with a structural coherence length of almost $10^4 \AA$.
The data are interesting for their similarity as well as their
differences from
the thermal conductivity of amorphous substances. We wish here to present a
model with which to interpret the experimental observations.

As is well-known, amorphous materials show a remarkably universal behavior
of the thermal conductivity $\kappa(T)$ at low temperatures \cite{note}.
At the lowest temperatures, around 1K, it has a power law
$\kappa(T) \sim T^2$. Around 10K it reaches a ``plateau" region where it
takes on a roughly constant value. This behavior is termed universal, as it
is observed in an extraordinarily diverse set of glassy materials -- from
polymer blends to vitreous silica to varnishes and resins. Chernikov et al
have obtained a similar $\kappa(T)$ curve for the icosahedral AlPdMn alloy :
it follows a power law  $T^\alpha$ with $\alpha \approx 2$ below 1.6K . It
then rises until it reaches the plateau region between
25K and 55K. The plateau is thus attained at a higher temperature than is
common in glasses, in addition, the value of $\kappa$ in the plateau region
is significantly higher as compared to the glasses.

The above observations have been made for temperatures that are around one-
tenth of the Debye temperature, which as deduced from specific heat data
is $\theta_D = 362K$. We may thus safely assume that the phonons -- or more
cautiously speaking, the vibrational modes -- responsible for thermal
conduction are of large spatial extent compared to interatomic distances. At
the very lowest temperatures, one may reasonably assume that
 phonons of long wavelength
propagate in the quasicrystal, as they do in disordered media, with
well-defined longitudinal and transverse sound velocities. For temperatures
below a couple of
Kelvin, thus, thermal transport takes place via these phonons, and $\kappa$
will depend on the scattering of the phonons by the structure. The
scattering mechanism the most frequently invoked in glasses is due to
localized ``tunneling systems" or TS \cite{tls}. In the AlPdMn quasicrystal,
low temperature sound velocity data \cite{bell} indicate the presence of
tunneling systems. Quasicrystals may have, in fact,
a natural candidate for TS in their so-called ``phason" defect states. These
are local atomic displacements that preserve allowed bond angles and
lengths.
The quasicrystal thus could scatter phonons
from these phason defects or from additional possible glassy-type
TS.

In the quasicrystal, we picture the large majority of the
eigenmodes of the perfect
quasiperiodic solid to be not phononlike, but rather modes with a
non-ballistic
type of propagation. One cannot speak of a dispersion relation in this case,
since there is no k-space, rather, we will speak of a mean square
displacement of the mode in real space as a function of time given by
$r^2(t) =\tilde D t^\alpha(\omega)$. Alternatively, one could write
an equivalent relation for the variation of the group velocity as a function
of energy(frequency). This dependence is empirically written, and
it is based on the hypothesis that eigenmodes on quasiperiodic structures are
 in between free-particle like ballistic modes and localized modes. This
hypothesis is primarily motivated by results of numerous studies of
tight-binding electronic models in quasicrystals (reviewed in \cite{fuj})
where a similar intermediate notion exists between extended and
localised electronic eigenfunctions -- the critical wavefunctions.
In consequence, electrons
have a dynamics in-between that of free electrons and completely
trapped ones. In addition to the indirect evidence for this coming
from the unusual scaling properties of the energy spectrum,
direct numerical evidence exists for
anomalous diffusion of particles in the quasiperiodic medium \cite{bps}.
In the present context of
the vibrational problem, Janssen and
Quilichini \cite{jq} have discussed numerical studies carried out for one,
two and three
dimensional models. The phonon spectra show the characteristic jagged
shape and multifractal scaling. As for the eigenfunctions, a numerical
calculation of the inverse participation ratio gives values that
are intermediate between that of localized states and extended states.
Although the values of $\alpha$ vary irregularly with $\omega$,
one can argue that the physics will be mainly determined by the underlying
smoothed out form
of the spectrum. We will therefore make the simplifying assumption that one
can
work with a locally averaged value of $\alpha$ which will be allowed to
vary smoothly with $\omega$ in the energy range of interest.
It is reasonable to suppose that this averaged value in
good quasicrystals is $2 > \alpha > \sim 1$ ($i.e.$ in the range given by
the low energy phonon value of 2, down to possibly
slower-than-normal-diffusion values). Finally, the constant
$\tilde D$
is an unknown parameter, to be eventually determined by comparison with
experiment.

In a
real sample, these modes will exist
from a length scale given by the coherence length $\xi_{coh}$, down to
atomic distances $a$. For the longer length scales, true phonons will be
present. In terms of the energy $\omega$ (or temperature), small $\omega$
or low temperature corresponds to exciting majoritarily phonons, while
higher temperatures will lead to the exciting of the quasilattice
vibrational modes -- we propose to term them ``quasons" from here on.
 Our model supposes that the density of states is that of phonons at very
low energies, crossing over to the quason form at an energy scale
$\omega_\xi$
that we can estimate using the values $\xi_{coh}/a \approx 10^{-3}$ and
$\theta_D $=362K. It corresponds to a crossover temperature of less than a
Kelvin. One should note, however, that no abrupt changes should be expected
at this crossover, since the low
energy
end of the quason spectrum is phononlike, with a
density of states not much different from the phonon one, until one
reaches somewhat higher temperatures.
For the range of temperature of a few Kelvin and below, we can thus
assume the density of states will be that of the phonons, while above the
crossover energy one will have the density of states of the quasicrystal:
\begin{eqnarray} \label{dos}
\nonumber
(\omega <\omega_\xi)  \qquad N(\omega) = {V\over (2 \pi^2 v_s^3)} \omega^2
\\ (\omega
>\omega_\xi) \qquad N(\omega) = {d_s V\over (a^3\omega_D^{d_s})}\omega^{(d_s
-1)}
\end{eqnarray}
 where $\omega_D = (6 \pi^2)^{(1/3)}
(\xi_{coh}/a)^{(3/d_s)}
(v_s/\xi_{coh})$ taking for illustration the simplest case of a unique value
of $d_s$ in the quason r\'egime. The spectral dimension
$d_s$ which is equal to 3 for phonons, is
$3\alpha/2$ for the quasons, which
 results in the density of
states in the quason-r\'egime growing somewhat more slowly with $\omega$
than for phonons.

At the very lowest temperatures, then, we may apply the usual formula giving
the phonon contribution to thermal conductivity,
\begin{equation} \label{pho}
\kappa(T) = {1\over 3} \int d\omega \quad v(\omega) C(\omega) l(\omega)
\end{equation}
where $C$ is the specific heat of modes of energy $\omega$:
$C(\omega) = N(\omega) \omega^2  e^{\beta
\omega}/((k_BT)^2(e^{\beta \omega}-1)^2)$.
The mean free path $l(\omega)$ is
determined by the scattering processes that enter in play. We now
discuss the very low and intermediate temperature
r\'egimes, by turn.

At the lowest temperatures, phonons carry the thermal current and the
TS referred to above give rise to $l \propto \omega^{-1}$ .
The specific heat in this limit is $C(\omega) \propto \omega^4
\exp(-\beta \omega)$. Setting the mode velocity $v(\omega)$ equal to
 the sound velocity, Eq.\ref{pho} readily yields \begin{equation}
\nonumber \kappa(T)= A T^2
\end{equation}
Chernikov et al have discussed the prefactor
$A$ as well as the exponent
of T in the above law assuming a plausible set of values for the parameters
entering in the explicit formula for $\kappa$. In fact, the
interpretation of the data is more complex than the simple argument
given above, and
we refer the reader to the original paper for a detailed discussion
of the behavior of $\kappa$ below 1.6K. We turn now to the expected
temperature dependence of
$\kappa$ at higher temperature.

At intermediate T, the relation in Eq.\ref{pho} is
replaced by an expression appropriate to the quason-dominated r\'egime
by
\begin{equation} \label{qua}
\kappa(T) = {\tilde D \over 3} \int d\omega \quad C(\omega) \tau^{\alpha -1}
(\omega)
\end{equation}
where we have introduced a scattering time $\tau$
for the
quason
modes. This scattering could be due to anharmonic interactions between
vibrational modes or scattering from defects in the structure. In glassy
compounds Rayleigh scattering from defects is often supposed, giving rise to
a frequency dependence $\tau(\omega) \sim \omega^{-4}$.
In vitreous silica,
Zeller and Pohl \cite{zp} have used the dominant phonon approximation along
with their thermal data to show a $\omega^{-4}$ dependence of
the scattering time $\tau$. This strong scattering at higher frequencies is
thought to be responsible for the failure of higher frequency phonons to
propagate, as they attain the Ioffe-Regel limit of $\l(\omega) \sim a$.
This leads to the levelling off of
the thermal conductivity at its ``plateau" value.

 In the quasicrystal alloy, lacking any
direct experimental evidence as to $\tau(\omega)$, one can try to apply
the equivalent
of the dominant phonon approximation. In this approximation, the thermal
conductivity is taken to be entirely determined by the modes which have
the highest heat capacity ($\omega_{dom}$ being given by the peak of
$C(\omega)$ for a given temperature).  Published
data on specific heat \cite{cbo2} gives a fit to a standard form for this
quantity at low temperature. Assuming a lattice contribution that is
similar to that in a glass and given the thermal
conductivity data, we conclude that the scattering time  $\tau$
has a similar $\omega$  dependence in the AlPdMn as in the glass.
system. This qualitative statement obviously needs further investigation.
although neutron scattering studies do see a significant
broadening effect away from the acoustic limit \cite{deb}, as we discuss
further below. This could imply that strong Rayleigh-type scattering is
found in the quasicrystal, just as in the amorphous case. The origin of this
scattering could be related to inhomogeneities of the structure as in the
glass. However in view of the good structural quality of the
quasicrystalline sample this seems an unsatisfactory explanation. A
second and interesting
possibility is that this scattering lifetime is an artifact arising from our
assumption that the density of states is a smoothly varying and $increasing$
function of $\omega$ in this energy range. In fact, it has been observed
in all the models studied
\cite{jq} that there are dips, or pseudogaps, in the density of states with
some important ones occurring as one approaches the middle of the spectrum.
If such a dip occurs at the energies just below the plateau energy, then one
may attribute the dependence of $\kappa$ in this region rather to a density
of states effect than to an increased scattering rate effect.
Neutron scattering \cite{deb} provides no evidence for such pseudogaps,
although one should note that we are speaking here of
longer length scales (several dozen $\AA$). There $is$ evidence from
neutron scattering for a significant broadening of modes away from the
acoustic limit. This favors the interpretation in terms of a lifetime
effect. In
any case, we will proceed on the hypothesis that it is the lifetime which is
strongly energy-dependent. The alternative scenario can be easily handled
by a similar type of calculation if necessary.

As in the amorphous case, the strong frequency dependence of $\tau$
then leads to the ``falling off" of $\kappa$(T) above a certain temperature.
Thus the
plateau appears for the quasicrystal, and for the same reason as in the
glasses.
The quantitative differences are important, as they arise from the
fundamental difference in the dynamics of the heat carrying modes in
these two systems. This is now discussed.
In this simple-minded model given by Eq.\ref{qua}
the thermal conductivity of quasicrystal and glass differ significantly
in that the scattering lifetime enters with an exposant $\alpha -1$ which is
in all cases smaller than one (which is the exponent for the glassy case).
If one takes the Rayleigh scattering $\tau$, then
\begin{equation} \label{final}
\kappa(T) \sim \int d\omega \quad C(\omega) \omega^{-4(\alpha -1)}
\end{equation}
Since the dominant frequency dependence is contained in the last factor,
which decreases less rapidly than for the usual phonon case,
the quason thermal conductivity is less strongly attenuated by the
Rayleigh scattering.
This implies that $\kappa$ will be slower in
reaching its plateau value, and
 explains the observed
shift of the plateau region to a higher temperature in AlPdMn. To make a
quantitative estimate of the shift expected one would require further
information on the prefactors entering in the scattering lifetime.

A related point concerns the relative magnitude of $\kappa$ in the
glassy and quasicrystalline systems. Chernikov et al have noted that
when data are plotted using
variables scaled by the measured values of the Debye temperature and sound
velocity, $\tilde \kappa (\tilde T)$
in the glass and the quasicrystal are very similar upto
 the glass plateau, beyond which the
quasicrystal $\tilde \kappa$ continues to rise to a higher value. This
could be understood as follows: in terms of appropriately scaled
quantities
the thermal capacity of phonons and phasons are roughly
equal -- this is indicated by the similar values of $\tilde \kappa(T)$
below the plateau for glass and quasicrystal, This would then imply
 since modes are slightly less strongly attenuated in the
quasicrystal, that
$\tilde \kappa$ continues to rise in the quasicrystal, attaining
a plateau value that is higher than for the glass. This is borne out by a
little model calculation (see figure) where we have integrated Eqs.\ref{qua}
and \ref{pho} numerically and compared the results for $\kappa(T)$. The
scattering time is taken to be identical for the two cases, namely an
interpolation between a constant value (at low frequency) and $\omega^{-4}$
(at higher frequency). The density of states is pure phonon for the lower
curve; for the upper curve, there is a crossover from phonon to quason
density of states (with $\alpha =3/2$).

 In
conclusion, we note that this discussion applies to the temperature r\'egime
$T< 50K$.
 The experimental data indicates an upturn of
$\kappa$ at the upper end of this range. As in the glasses, this
increase of the thermal conductivity could be explained by at least two
scenarios.  One is a scenario in which the Rayleigh-scattering is somehow
``turned off" at higher frequencies, so that modes can propagate anew in the
sample. The other would be a new type of heat conduction in which
localized vibrational modes begin to hop, aided by anharmonic
interactions with ordinary phonons \cite{ray}. However this high
temperature r\'egime is outside the scope of our present paper.

\bibliographystyle{prsty}

\begin{thebibliography}{12}
\bibitem{note}
It
should be understood that the thermal conductivity in question is that due
to the lattice, $i.e.$ after subtraction of the electronic part if any.
This introduces an uncertainty in what value to assign to each contribution
at very low temperatures, but at the temperatures of interest to us here,
the electronic contribution is negligibly small compared to the phonon
contribution.
\end{thebibliography}
\newpage
FIGURE CAPTION

Results for $\kappa$ (arbitrary units) vs T from simple model calculation
(see text). The upper curve corresponds to the quasicrystal, the lower curve
to a glass having the same Debye temperature and scattering rate
$\tau(\omega)$.

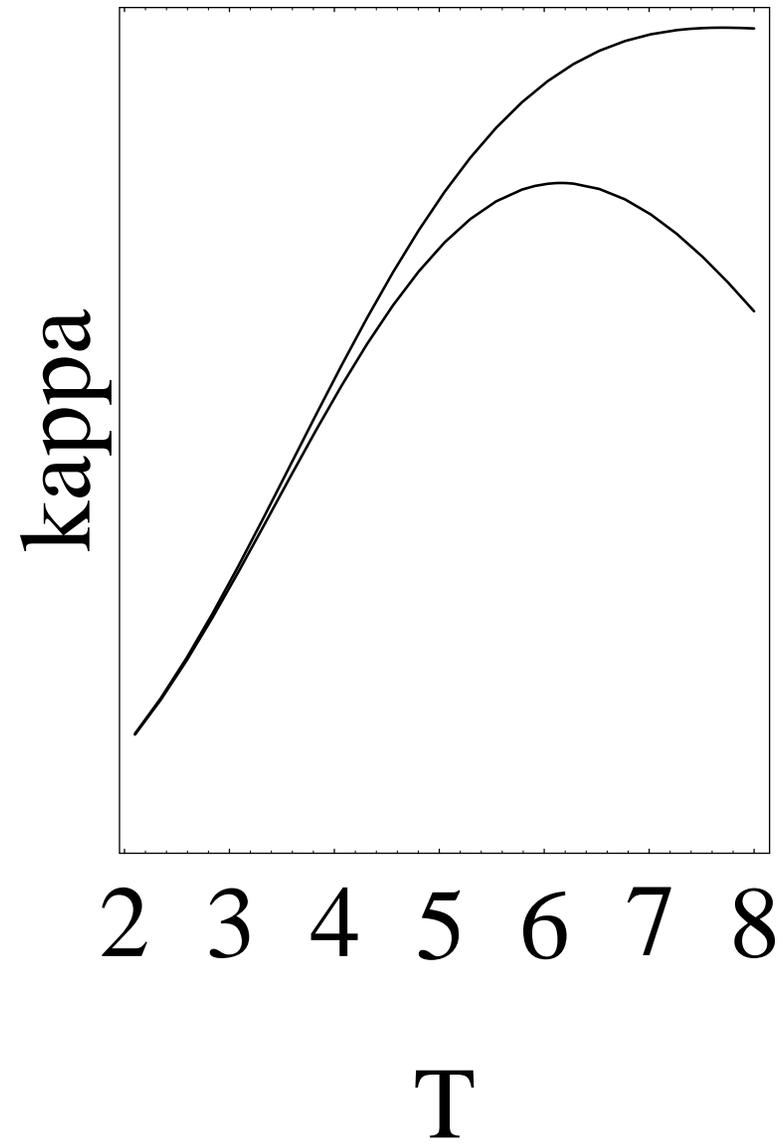
\end{document}